\documentclass[twocolumn]{aastex631}

\shorttitle{GJ 357 d: Potentially Habitable World or Agent of Chaos?}
\shortauthors{Stephen R. Kane \& Tara Fetherolf}

\begin{document}

\title{GJ 357 d: Potentially Habitable World or Agent of Chaos?}

\author[0000-0002-7084-0529]{Stephen R. Kane}
\affiliation{Department of Earth and Planetary Sciences, University of
  California, Riverside, CA 92521, USA}
\email{skane@ucr.edu}

\author[0000-0002-3551-279X]{Tara Fetherolf}
\affiliation{Department of Earth and Planetary Sciences, University of
  California, Riverside, CA 92521, USA}


\begin{abstract}

Multi-planet systems provide important laboratories for exploring
dynamical interactions within the range of known exoplanetary system
architectures. One such system is GJ 357, consisting of a low-mass
host star and three orbiting planets, the outermost (planet d) of
which does not transit but lies within the Habitable Zone (HZ) of the
host star. The minimum mass of planet d causes its nature to be
unknown, both in terms of whether it is truly terrestrial and if it is
a candidate for harboring surface liquid water. Here, we use three
sectors of photometry from the Transiting Exoplanet Survey Satellite
(TESS) to show that planets c and d do not transit the host star, and
therefore may have masses higher than the derived minimum masses. We
present the results for a suite of dynamical simulations that inject
an Earth-mass planet within the HZ of the system for three different
orbital and mass configurations of planet d. These results show that
planet d, rather than being a potentially habitable planet, is likely
a source of significant orbital instability for other potential
terrestrial planets within the HZ. We find that relatively small
eccentricities of planet d cause a majority of the HZ to be unstable
for an Earth-mass planet. These results highlight the importance of
dynamical stability for systems that are prioritized in the context of
planetary habitability.

\end{abstract}



\section{Introduction}
\label{intro}

The vast number of exoplanet discoveries have allowed the statistical
analysis of planetary systems, inferences of planetary demographics,
and their relationship to formation and evolution processes. The
orbital and mass distribution within planetary systems has yielded
significant insight into the nature of planetary architectures
\citep{ford2014,winn2015,mishra2023a,mishra2023b} and their
relationship to the layout of the solar system
\citep{horner2020b,kane2021d}. Likewise, the diversity of planetary
masses detected has enabled the investigation of the transition
between terrestrial bodies and planets with a far more substantial
gaseous envelope
\citep{weiss2014,rogers2015a,wolfgang2016,chen2017,unterborn2023}. For
those planets within the terrestrial regime, the highest priority
targets for further study are frequently those that lie within the
Habitable Zone (HZ) of the host star
\citep{kasting1993a,kane2012a,kopparapu2013a,kopparapu2014,kane2016c,hill2018,hill2023}. However,
presence within the HZ is not a sufficient requirement for planetary
habitability as there are a vast number of stellar, planetary, and
system properties that can influence the long-term sustaining of
surface liquid water. Among these many habitability factors are the
eccentricity
\citep{williams2002,dressing2010,kane2012e,linsenmeier2015,kane2017d}
and dynamical stability
\citep{kopparapu2010,kane2015b,kane2019e,kane2022a} of planets within
the HZ, both of which can play a crucial role in their insolation flux
variability, or even orbital viability.

A planetary system of recent interest is the GJ~357 system, with an
architecture consisting of an M dwarf star harboring three known
planets with orbital periods of 4, 9 and 56 days. The planetary system
was initially detected via photometry from the Transiting Exoplanet
Survey Satellite (TESS) \citep{ricker2015,guerrero2021a,kane2021b} as
the innermost planet was observed to transit the host star. The two
outer planets in the system were subsequently detected through radial
velocity (RV) observations and reported by two separate teams:
\citet{jenkins2019b} and \citet{luque2019b}. The RV measurements
provided a mass for the transiting inner planet and facilitated
atmospheric loss models for the planet in combination with X-ray
observations of the host star \citep{modirroustagalian2020c}. However,
no transits had been detected for the outer two planets and so their
measured masses of 3.4~$M_\oplus$ for the middle planet and
6.1~$M_\oplus$ for the outer planet (planet d) were stated as minimum
masses \citep{luque2019b}, leaving open the possibility that
subsequent TESS data may yet reveal their transits. Planet d was cited
as being of particular interest since it lies within the HZ of the
host star, and \citet{kaltenegger2019a} discussed in detail the
potential climate in the context of planetary habitability and
pathways toward observational confirmation. These discussions assumed
the RV mass to be the true mass of the planet, that the planet is
terrestrial in nature, and that it lies in a circular orbit.

In this paper, we present new data and calculations for the GJ 357
system to assess the effect that planet d has on the
HZ. Section~\ref{arch} provides calculations of the system HZ, and a
discussion of the possible terrestrial nature of planet d, the
eccentricity of the orbit, and the detectability of other terrestrial
planets in the system. We also present new TESS photometry that rules
out transits for planets c and d, and discuss the implications for
their true masses. In Section~\ref{dynamics}, we describe our
dynamical simulation that assesses the dynamical viability of an
Earth-mass planet within the HZ in the presence of planet d for three
different configurations of the planet d mass and eccentricity. We
also examine individual cases of injected planets that survive the
simulation, but are not long-term stable. Section~\ref{discussion}
discusses the consequences of these results for long-term system
stability of possible habitable planets, and the implications for
exoplanet demographics within the HZ. We provide suggestions for
further work and concluding remarks in Section~\ref{conclusions}.


\section{System Architecture}
\label{arch}

Here we describe the architecture of the system, calculate the extent
of the HZ, and examine TESS data in the context of additional
planetary transits.


\subsection{Orbits and Habitable Zone}
\label{orbits}

\begin{figure}
  \includegraphics[width=8.5cm]{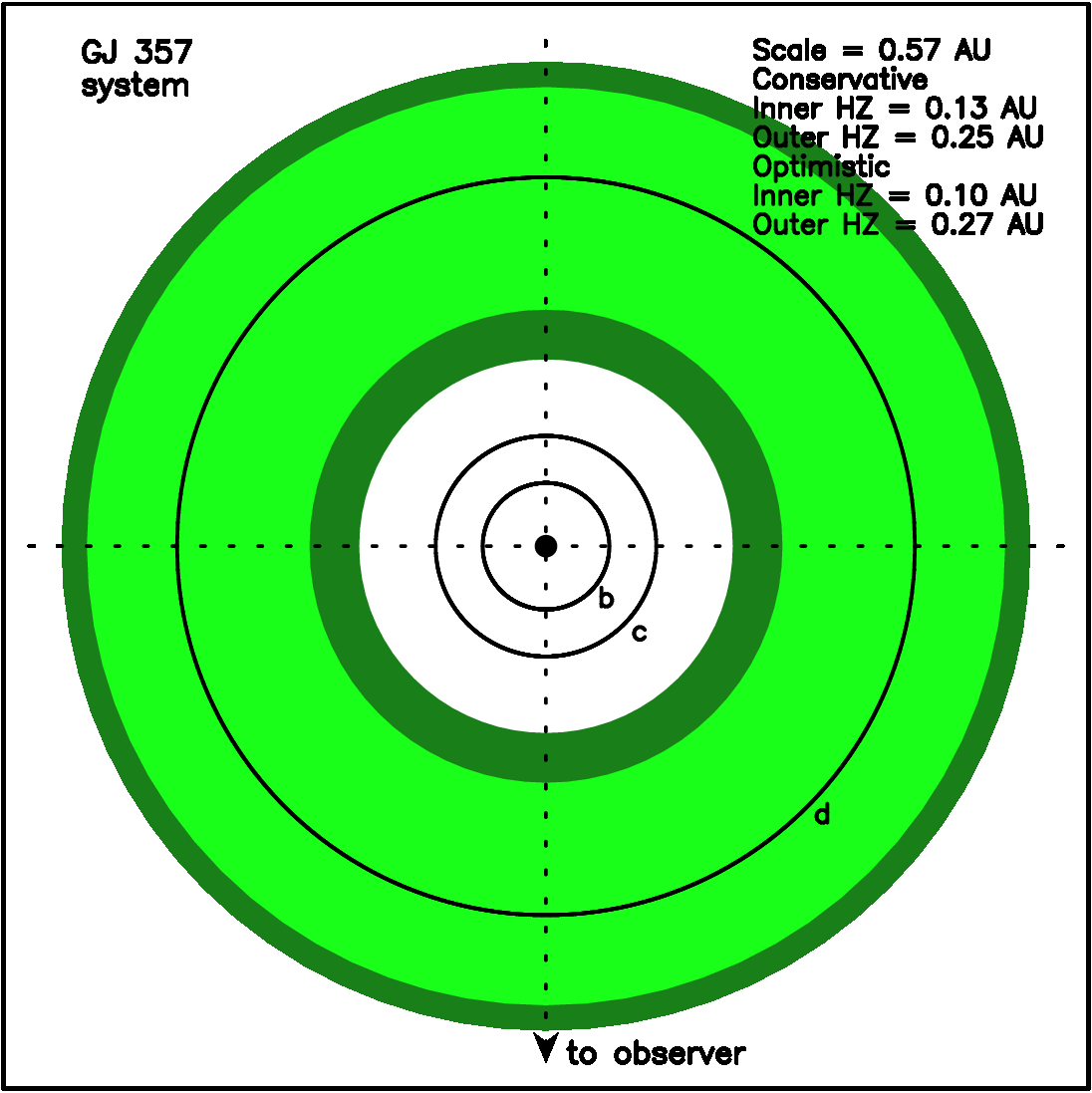}
  \caption{The architecture and HZ of the GJ~357 system, where the
    scale of the figure is 0.57~AU along each side. The CHZ is shown
    in light green, the OHZ extensions to the HZ are shown in dark
    green, and the orbits of the known planets are shown as solid
    circles.}
  \label{fig:hz}
\end{figure}

As described in Section~\ref{intro}, the GJ~357 system consists of
three known planets orbiting a low-mass star. For the analysis in this
work, we adopt the stellar and planetary parameters of
\citet{luque2019b}. The host star has a spectral classification of
M2.5V, with a mass of $M_\star = 0.342$~$M_\odot$, a radius of
$R_\star = 0.337$~$R_\odot$, an effective temperature of
$T_\mathrm{eff} = 3505$~K, and a luminosity of $L_\star =
0.01591$~$L_\odot$. These properties allow us to calculate the HZ of
the system, including the conservative HZ (CHZ) and the optimistic
extension to the HZ (OHZ) based upon the assumption that Venus and
Mars had surface liquid water in their past, described in detail by
\citet{kane2016c}. We calculate distance ranges of 0.131--0.254~AU and
0.103--0.268~AU for the CHZ and OHZ, respectively. The extent of the
HZ and the orbits of the known planets are shown in
Figure~\ref{fig:hz}, where the CHZ is shown in light green and the OHZ
is shown in dark green, and the semi-major axes of the planetary
orbits are 0.35~AU, 0.061~AU, and 0.204~AU for the b, c, and d
planets, respectively. It is worth noting that \citet{jenkins2019b}
refer to the innermost planets of the system as ``c'' and ``b'' in
order of increasing semi-major axis, whereas \citet{luque2019b} refer
to those same planets as ``b'' and ``c''. Here, we adopt the naming
convention of \citet{luque2019b}, and thus the three planets are
referred to as ``b'', ``c'', and ``d'' in order of increasing
semi-major axis.

\begin{figure*}
  \begin{center}
    \includegraphics[width=16.0cm]{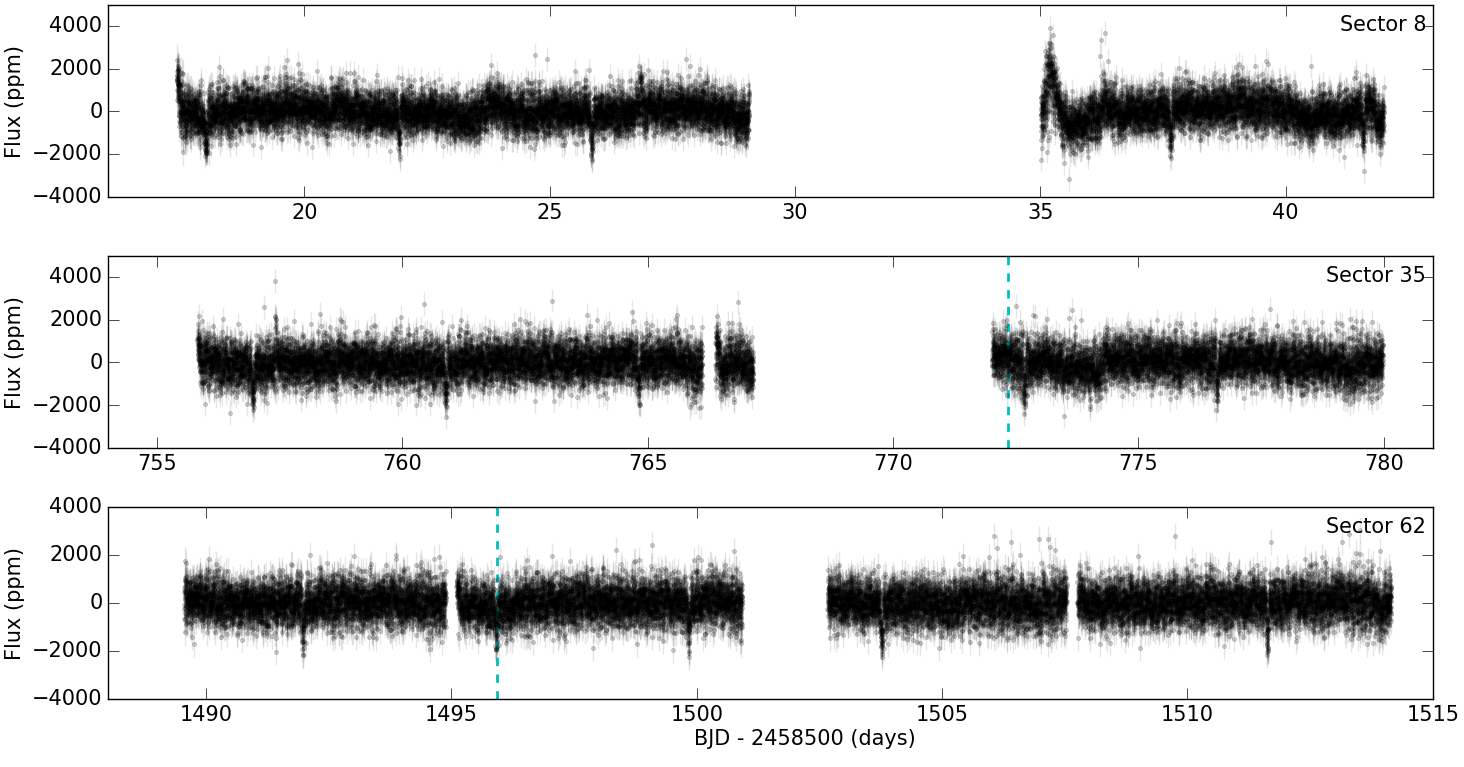}
  \end{center}
  \caption{The TESS PDCSAP light curve of GJ~357, which includes
    photometry obtained during Sectors 8 (top panel), 35 (middle
    panel), and 62 (bottom panel). Transits of the inner planet,
    GJ~357~b, can be seen in the light curve every $\sim$4 days. The
    vertical dashed lines in the middle and bottom panels indicate the
    predicted inferior conjunction passage for planet d.}
  \label{fig:tess}
\end{figure*}

There are various components of the GJ~357 system that remain
unconstrained. The planetary orbits shown in Figure~\ref{fig:hz} are
assumed to be circular, and indeed statistical studies have found that
smaller planets tend to have relatively low eccentricities
\citep{kane2012d,vaneylen2015}. \citet{luque2019b} considered
eccentric orbits in the preliminary analysis of the RV data, but
assumed circular orbits for their final model due to comparable fits
between eccentric and non-eccentric cases, and the computational
burden of including eccentricity as a free
parameter. \citet{jenkins2019b} also explored non-zero eccentricities,
which included a 1$\sigma$ eccentricity for planet d consistent with
$\sim$0.1, but fixed circular orbits in the final analysis. Thus, an
eccentric orbit for planet d is allowable with the present data for
the system.

The true architecture of the system is only known to the extent that
the sensitivity of the observational data allows, including whether
there may be further planets within the system. For example, an
additional planet of Earth-mass that lies within the HZ would be
challenging to detect with the present dataset. We calculate that an
Earth-mass planet at the inner and outer edges of the OHZ would have
RV semi-amplitudes of 0.47~m/s and 0.29~m/s, respectively, which fall
below the 2~m/s rms precision of the utilized spectrographs. We
therefore find that an additional Earth-mass planet in the HZ would
likely remain undetectable with the current RV data.

According to \citet{luque2019b}, planets c and d have minimum masses
of 3.4 and 6.1 Earth masses, respectively. Since planets c and d are
not presently known to transit, the true mass of the planets may be
significantly higher than the minimum masses. A dynamical analysis of
the system performed by \citet{luque2019b} did not place significant
constraints on the inclination of planets c and d. However, the
dynamical analysis performed by \citet{jenkins2019b}, whose RV
analysis derived slightly higher planetary masses than those found by
\citet{luque2019b}, determined that the mass of planet d lies in the
range of 7.2--11.2 Earth masses. The resolution on whether or not
planet d transits the host star is an important component of
understanding the true nature of this HZ planet.


\subsection{Planet d Does Not Transit}
\label{notransit}

The analysis performed by \citet{jenkins2019b} and \citet{luque2019b}
used Sector 8 of TESS photometry, which occurred during the TESS Prime
Mission. Since then, GJ~357 has been observed again during sectors 35
and 62. The time-series photometry observed by TESS was obtained
through the Mikulski Archive for Space Telescopes (MAST; \dataset[DOI:
  10.17909/t9-nmc8-f686]{http://dx.doi.org/10.17909/t9-nmc8-f686}). We
utilize the 2-min pre-search data conditioning simple aperture
photometry (PDCSAP), which was processed by the Science Processing
Operations Center (SPOC) pipeline
\citep{jenkins2016}. Figure~\ref{fig:tess} shows the light curve of
GJ~357 for all three TESS sectors, where the transits of GJ~357~b can
be plainly seen in the light curve every $\sim$4 days. The transit
depth is 1095~ppm, which translates to a radius for planet b of $R_p =
1.217$~$R_\oplus$. Given that the average rms scatter for the shown
three TESS sectors is 620~ppm, and that planets c and d are more
massive than planet b, transits of the outer two planets should be
visible in the data if they occur. Indeed, planets of size
1.5~$R_\oplus$ and 2.0~$R_\oplus$ would produce transit depths of
1665~ppm and 2960~ppm, respectively. Assuming the lowest of these
radii values, transits of planets c and d are ruled out at $2.7\sigma$
per TESS measurement, which is equivalent to that expected for a
grazing transit. Based on the system properties described in
Section~\ref{orbits}, we estimate central transit durations of
$\sim$2.0 and $\sim$3.5 hours for planets c and d, respectively. Thus,
central transits for planets c and d are ruled out at a significance
of $\sim$$21\sigma$ and $\sim$$28\sigma$, respectively.

Given the expected transit depths and photometric precision, the three
TESS Sectors shown in Figure~\ref{fig:tess} are sufficient to confirm
that planet c (whose orbital period is 9.12~days) does not
transit. Planet d has an orbital period of 55.661~days, and so its
alignment with a particular TESS sector is less obvious than for
planet c. However, the RV orbit shows that an inferior conjunction
passage missed Sector 8 entirely, as noted by both
\citet{jenkins2019b} and \citet{luque2019b}. Using the planet d
orbital ephemeris of \citet{luque2019b}, we calculate BJD times of
inferior conjunction of 2459272.34 and 2459995.93, which occur during
the sector 35 and 62 observing windows, respectively. These inferior
conjunction times are indicated by the vertical dashed lines shown in
Figure~\ref{fig:tess}. The inferior conjunction for sector 35 occurs
directly after the telemetry data gap, and reveals no evidence for a
transit. The inferior conjunction for sector 62 also falls within the
TESS photometry, and in fact falls at the same time as a planet b
transit, which would have resulted in a syzygy transit event had
planet d transited \citep{luger2017c,veras2017d}. The lack of
observable signature at either location verifies that planet d does
not transit the host star.

As noted in Section~\ref{orbits}, the confirmation that planet d does
not transit means that the planetary mass may be significantly higher
than the minimum mass of 6.1~$M_\oplus$. There have been numerous
derivations of mass-radius relationships for exoplanets that estimate
the upper limits for a terrestrial body
\citep{dressing2015a,rogers2015a,chen2017}. The empirical relationship
between planet mass and radius derived by \citet{chen2017} detected a
transition from terrestrial into ``Neptunian'' planets with a greater
volatile inventory at a boundary of $\sim$2~$M_\oplus$. Without a
radius measurement, there is a great deal of degeneracy regarding the
bulk properties of a planet that possibly lies within the terrestrial
regime, even if the true mass of the planet is known
\citep{valencia2007b,dorn2015,zeng2016a}. The sub-solar metallicity of
GJ~357 \citep{luque2019b}, combined with the relatively high mass of
planet d, suggests the planet lies outside the nominal rocky planet
zone \citep{unterborn2023}. Furthermore, \citet{kopparapu2014}, who
assume that planets with masses larger than 5~$M_\oplus$ are not
rocky, found an increasing rate for the outgoing longwave radiation
with planet mass due to the smaller atmospheric column depth,
decreasing greenhouse warming and increasing the width of the HZ at
the inner edge. With all of these considerations in mind, it is
difficult to state with any certainty what the true nature of planet d
is, but the minimum planetary mass allows for a large parameter space
where a habitable scenario is increasingly unlikely.


\section{Dynamical Stability Within the Habitable Zone}
\label{dynamics}

Here we provide the details and results of an investigation into the
dynamical viability of planetary orbits throughout the HZ of the
GJ~357 system in the presence of the three known planets.


\subsection{Simulation Description}
\label{setup}

We adopt the methodology described by \citet{kane2019c,kane2021a}, in
which the Mercury Integrator Package \citep{chambers1999} was applied
with a hybrid symplectic/Bulirsch-Stoer integrator with a Jacobi
coordinate system \citep{wisdom1991,wisdom2006b}. Each simulation was
integrated for $10^6$ years, equivalent to $\sim$$6.5 \times 10^6$
orbits of planet d, and with a time step of 0.1~days to ensure
adequate resolution of planet-planet encounters that involve the
innermost planet. We used the stellar and planetary properties
provided by \citet{luque2019b} and the HZ boundaries calculated in
Section~\ref{orbits}. We tested the orbital stability within the HZ by
injecting an Earth-mass planet in a circular orbit that is coplanar
with planets b and c, the latter of which is assumed to have a
near-edge on orbit that allows the minimum planet mass to be adopted
as an approximation of the true mass. The new planet was injected at
semi-major axes within the range 0.1--0.27~AU and in steps of 0.001,
encompassing the full OHZ range of 0.103--0.268~AU. Additionally, each
semi-major axis step incorporated initial evenly-spaced mean anomalies
of 60\degr, 180\degr, and 300\degr for the injected planet.

Because the orbit of planet d is poorly constrained, our simulations
were conducted for three specific orbital configurations of planet
d. Firstly, we considered the case of planet d being near coplanar
with the other planets such that, like planet c, the minimum planet
mass (6.1~$M_\oplus$) is a reasonable approximation of the true
mass. Secondly, we considered the case of the planet d orbit being
significantly inclined with respect to the orbital plane of the other
planets, producing a planet mass of 10.0~$M_\oplus$, which is within
the dynamical limits found by \citet{jenkins2019b}. Thirdly, we
considered the case where the orbital inclination and planet mass of
planet d is the same as the second case, but now with a slight
eccentricity of 0.1. The combination of these three cases resulted in
$\sim$1500 simulations in total.


\subsection{Simulation Survival Rates}
\label{survival}

\begin{figure*}
  \begin{center}
    \includegraphics[width=16.0cm]{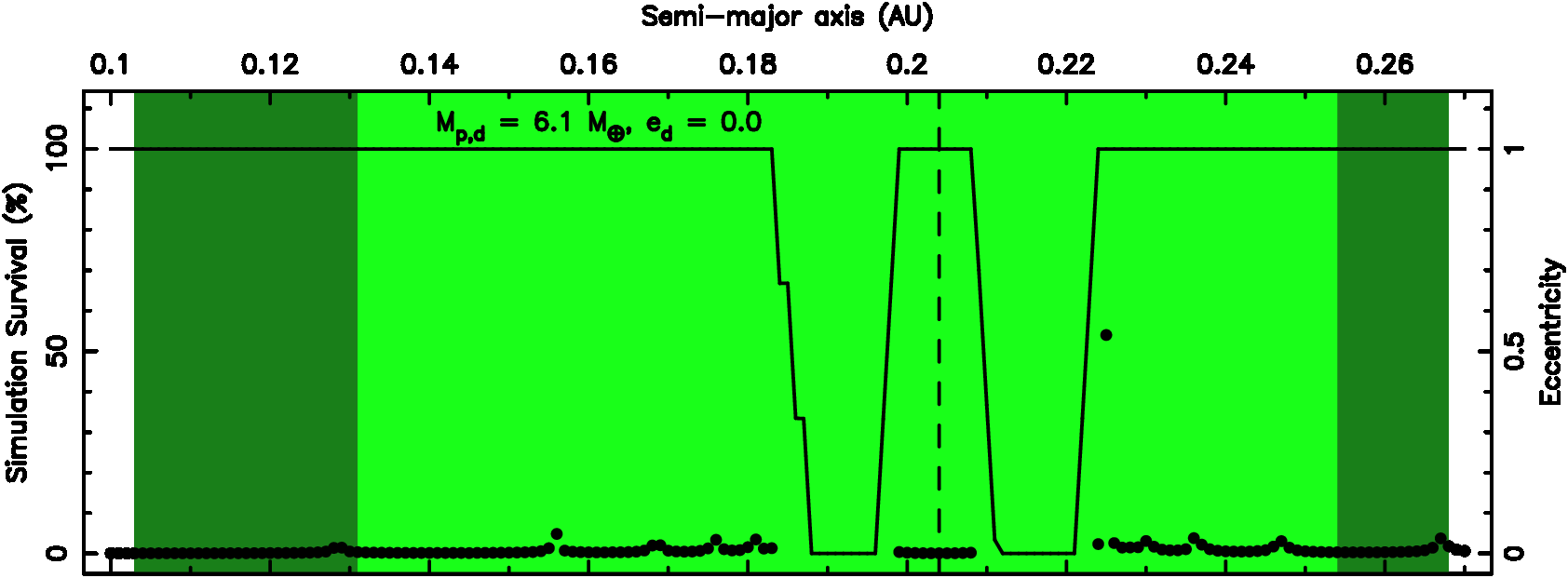} \\
    \includegraphics[width=16.0cm]{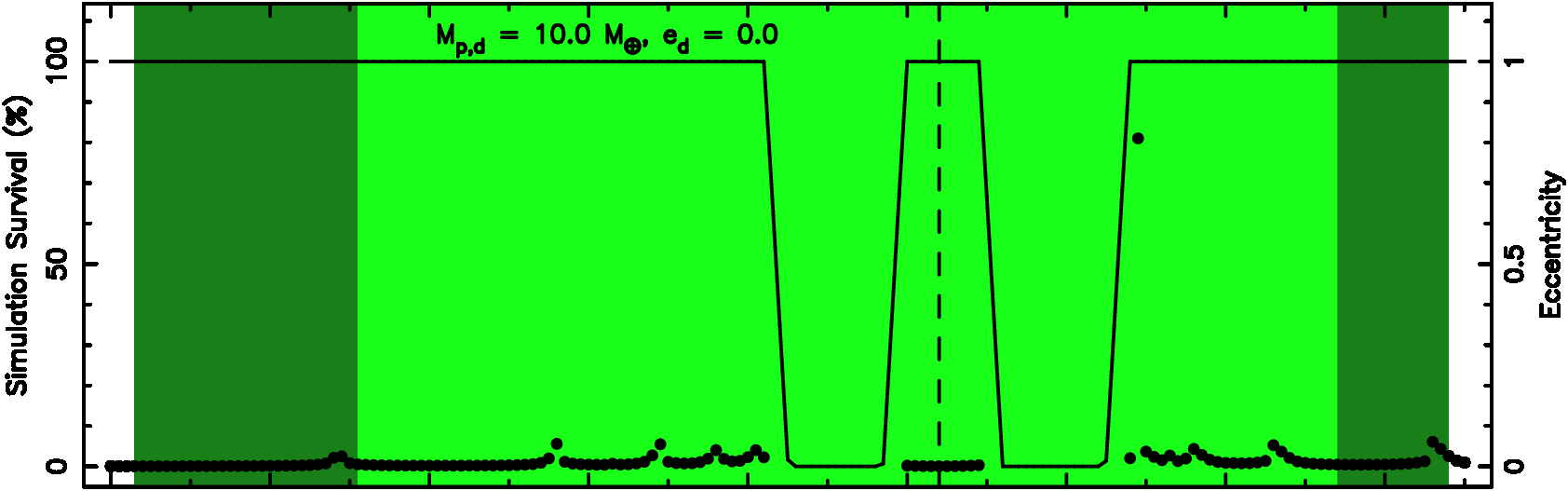} \\
    \includegraphics[width=16.0cm]{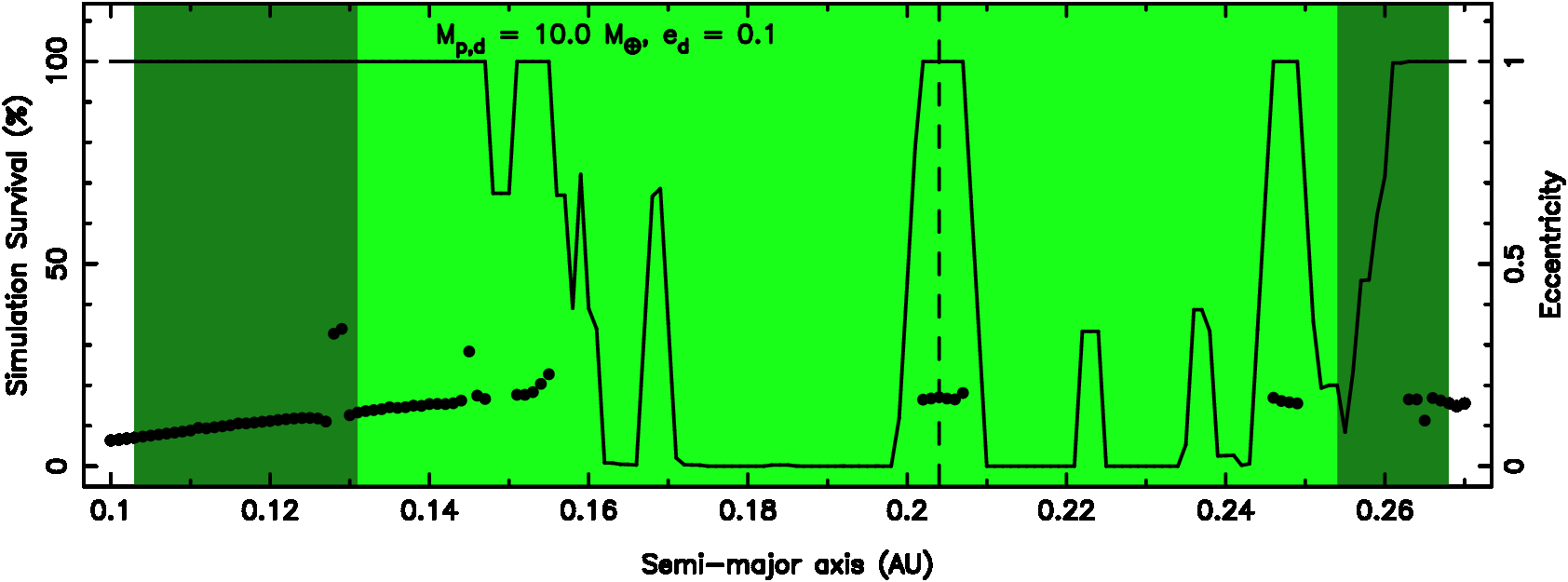}
  \end{center}
  \caption{Dynamical stability results for the system architecture
    cases that vary the inclination, mass, and eccentricity of planet
    d. Each panel shows the percentage of the simulation that the
    injected planet survived as a function of semi-major axis, shown
    as a solid line. As for Figure~\ref{fig:hz}, the CHZ is shown in
    light green and the OHZ is shown in dark green. The vertical
    dashed line indicates the semi-major axis of planet d, and the
    mass and eccentricity for planet d are labeled near the top of
    each panel. The maximum eccentricities of the injected planet
    through the simulations are shown as black dots (see
    Section~\ref{ecc}).}
  \label{fig:sim}
\end{figure*}

The outcome for each of the simulations described in
Section~\ref{setup} was assessed based on the survival of the injected
planet, where non-survival can mean the planet was either ejected from
the system or lost to the gravitational well of the host star. Note
that the Mercury Integrator Package assumes point-masses, and so
planet-planet collisions, though possible, are not considered in the
simulations. The results of the simulations are summarized in
Figure~\ref{fig:sim}, which contains three panels that represent the
results for each of the three orbital configuration cases for planet d
mass and eccentricity, shown near the top of each panel. Each panel
shows as a solid line the survival time of the injected planet (as a
percentage of the full $10^6$~year integration) as a function of the
semi-major axis of the injected planet. The HZ is depicted as for
Figure~\ref{fig:hz}, with the CHZ shown in light green and the OHZ
shown in dark green. The semi-major axis of planet d is indicated by
the vertical dashed line. We detected no cases in which any of the
three known planets did not survive the simulations, largely due to
their substantial mass compared with the injected planet.

For all three cases, there is an island of stability around the
semi-major axis of planet d, allowing for the possibility of Trojan
planets in similar orbits \citep{paez2015a}. Remarkably, such Trojan
planetary orbits can maintain long-term stability
\citep{cresswell2009,schwarz2009a}, although eccentricity of the
primary planet reduces this stable region \citep{dvorak2004}, as seen
in the bottom panel of Figure~\ref{fig:sim}. For the first case
($M_{p,d} = 6.1$~$M_\oplus$ and $e_d = 0.0$; top panel), the presence
of planet d clears a substantial region around the orbit, and 18\% of
the HZ is rendered unstable. For the second case ($M_{p,d} =
10.0$~$M_\oplus$ and $e_d = 0.0$; middle panel), the instability
regions surrounding the orbit of planet d slightly expand to occupy
21\% of the HZ. For the third case ($M_{p,d} = 10.0$~$M_\oplus$ and
$e_d = 0.1$; bottom panel), the introduction of a relatively small
eccentricity to the planet d orbit greatly increases the instability
with the HZ, resulting in 60\% of the HZ being unstable for the
injected planet. Instability of the injected planet primarily arises
though gravitational perturbations from the known planets,
particularly at mea motion resonance (MMR) locations, that increase
the eccentricity of the injected planet. Such eccentricity increases
are often lead to more frequent perturbations that culminate in the
ejection of the planet from the system. These simulation results show
that, for the planet mass range explored, increasing the eccentricity
of planet d has a larger effect on the instability within the HZ than
increasing the planet mass. However, regions of the HZ where the
injected planet remains in the system does not guarantee that the
planet's orbit is conducive toward potential habitability.


\subsection{Eccentricity Consequences}
\label{ecc}

If the injected terrestrial planet in our simulations survived the
$10^6$ year integration time, there may remain orbital consequences
from interacting with the other planets in the system. In general,
compact planetary architectures benefit from stability enabled by the
planets' relatively small Hill radii, since that scales linearly with
semi-major axis. The results shown in Section~\ref{survival}
demonstrate that increasing planet mass, and therefore Hill radius,
gradually increases the region of instability surrounding the
planet. However, increasing eccentricity has a far greater effect on
system dynamics through conservation of angular momentum, and even
stable configurations can inherit significant eccentricity evolution
cycles.

\begin{figure*}
  \begin{center}
    \includegraphics[width=16.0cm]{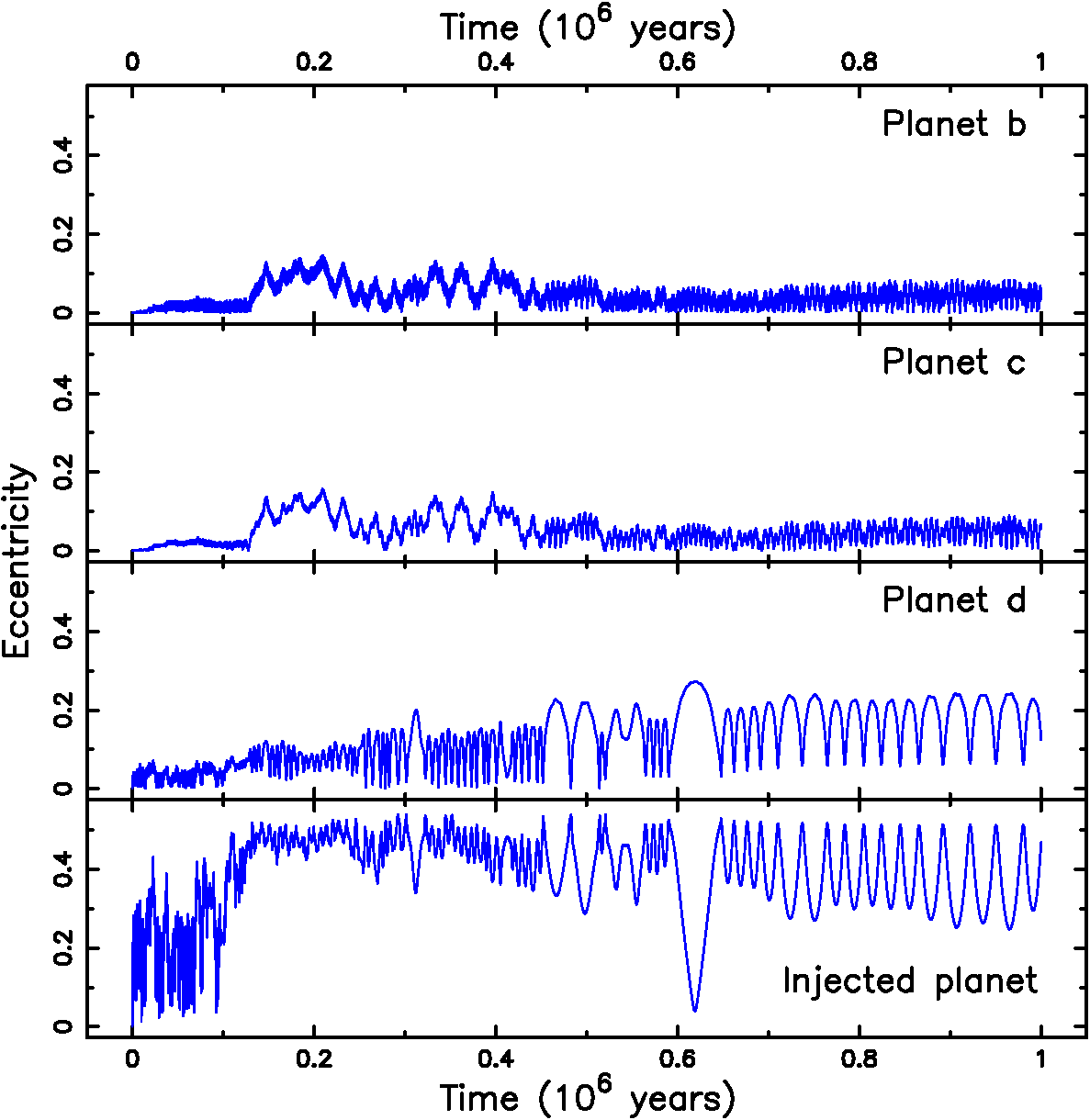}
  \end{center}
  \caption{Eccentricity evolution over $10^6$~years for the three
    known GJ~357 planets and the injected planet for the case of
    $M_{p,d} = 6.1$~$M_\oplus$ and $e_d = 0.0$ and the inserted planet
    at a semi-major axis of 0.225~AU.}
  \label{fig:ecc}
\end{figure*}

Another feature shown in Figure~\ref{fig:sim} is the maximum
eccentricity achieved by the injected planet, indicated as black dots
in each of the panels. These are shown for the cases where the planet
survives the full integration time. The maximum eccentricity values
shown for the first two architecture cases (top and middle panels)
show that the injected planet eccentricities remain low when the
initial conditions of planet d assume a circular orbit. Exceptions to
this include slight eccentricity increases at locations of MMR, such
as 0.129~AU and 0.155~AU, corresponding to 2:1 and 3:2 MMR with planet
d, respectively. More significant exceptions are those close to the
instability regions surrounding planet d, where planetary orbits lie
at the edge of chaotic instability. An example of this is shown in
Figure~\ref{fig:ecc}, which provides the eccentricity evolution for
all four planets in the system in the case where $M_{p,d} =
6.1$~$M_\oplus$, $e_d = 0.0$, and the injected planet has a semi-major
axis of 0.225~AU. The data shown in Figure~\ref{fig:ecc} are the first
$10^6$~years of an extended $10^7$~year integration conducted to
explore the longer-term stability for this particular
architecture. The interaction with planet d quickly raises the
eccentricity of the injected planet into a quasi-chaotic state where
it remains up until $\sim$$0.5 \times 10^6$~years, even influencing
the eccentricity evolution of planets b and c. Beyond $\sim$$0.5
\times 10^6$~years, the injected planet enters into a stable periodic
exchange of angular momentum with planet d. Though the injected planet
survives the $10^7$~year integration, there is no guarantee that the
system will retain stability beyond the simulated period.

The third architecture case (bottom panel of Figure~\ref{fig:sim})
shows a much larger excitation of the injected planet eccentricities,
where the maximum eccentricity increases with increasing semi-major
axis and remains at $\sim$0.2 for the majority of 100\% survival
simulations. We conducted $10^7$~year simulations for several initial
semi-major axis values of the injected planet and found that the
planet either did not survive the full simulation or showed increasing
signs of chaotic behavior until the end of the simulation. Thus, many
of the injected planets in the third architecture case are unlikely to
maintain long-term stability beyond the time frame of the simulations
shown in Figure~\ref{fig:sim}, particularly at the locations of MMR,
due to the chaotic nature of the induced eccentric orbits.


\section{Discussion}
\label{discussion}

\citet{jenkins2019b} conducted long-term stability analyses of the
GJ~357 system and established that their provided orbital parameters
provide a stable configuration. The simulations described in
Section~\ref{dynamics} were integrated for $10^6$ years and, based on
the number of orbits for planet d within that time (see
Section~\ref{setup}), are generally sufficient to explore the
dynamical stability of the presented architecture cases. However,
based on the eccentricity evolution results reported in
Section~\ref{ecc}, many of the simulations show evidence for a chaotic
divergence from stable configurations beyond the $10^6$ time window,
which is consistent with the dynamical results for compact
multi-planet systems provided by \citet{tamayo2020b}. There are
numerous circumstances whereby compact multi-planet systems may
exhibit the onset of chaotic orbits, most particularly locations of
MMR and subsequent secular evolution that modulates MMR widths
\citep{tamayo2021a,tamayo2021b}, such as that seen in
Figure~\ref{fig:ecc}. Indeed, even the inner planets of the solar
system exhibit chaotic behavior over sufficiently long timescales
\citep{laskar1994,laskar1996b}. Therefore, the width of the
instability regions described in Section~\ref{dynamics} and shown in
Figure~\ref{fig:sim} may be considered a lower limit on the induced
instability by planet d for each of the three architecture cases.

Terrestrial planets may be optimally packed within the HZ for a wide
range of spectral types, provided the orbits are sufficiently circular
\citep{obertas2017,kane2020b}. As mentioned in Section~\ref{intro},
there has been previous consideration of orbital eccentricity effects
on planetary habitability
\citep{williams2002,dressing2010,kane2012e,linsenmeier2015,kane2017d}. Specifically,
the orbital modulation of stellar flux received at the top of the
atmosphere will influence the planetary climate, depending on the
eccentricity and the thermal inertia of the atmosphere that determines
the radiative equilibrium timescale
\citep{iro2010,way2017a,kane2021a}. Furthermore, eccentric orbits may
induce pseudo-synchronous spins states \citep{dobrovolskis2007b} and
obliquity variations \citep{deitrick2018a,deitrick2018b,vervoort2022}
that impact seasonal modulation of the planetary climate. Planet d
will always exchange angular momentum with an injected planet in the
HZ, leading to increased eccentricity, often to the point of
ejection. Under these conditions, harboring a stable habitable planet
within the HZ of GJ~357 system is therefore a challenging scenario.

GJ~357~d may not be a habitable planet, or even terrestrial, and may
act to exclude other potentially habitable planets from being present
in the system. On the other hand, planet c is almost half the minimum
mass of planet d, and is thus more likely to be terrestrial in
nature. According to \citet{luque2019b}, planet c received a factor of
4.45 more flux from the host star than Earth receives from the
Sun. Since planet c lies in the Venus Zone
\citep{kane2014e,vidaurri2022b}, it may be an excellent candidate for
a super-Venus, a phrase first coined for the planet Kepler-69c
\citep{kane2013d}. As a non-transiting planet, characterizing the
atmosphere of planet c requires facing the challenge of measuring
infrared excess \citep{stevenson2020c}, which may be achievable with a
space-based mid-infrared low-resolution spectrograph
\citep{mandell2022}. Moreover, GJ~357~b, with a mass of $M_p =
1.84$~$M_\oplus$, radius of $R_p = 1.217$~$R_\oplus$, and an incident
flux of 12.6 times the solar constant \citep{luque2019b} is another
interesting Venus analog candidate that was identified as such by
\citet{ostberg2023}. Given the chaotic orbital dynamics resulting from
planet d, and the potential for the other known planets to be Venus
analogs, the true value of the GJ~357 system may be realized in
exploring the boundaries of planetary habitability rather than
habitable environments.


\section{Conclusions}
\label{conclusions}

The GJ~357 system is a fascinating addition to the rapidly growing
demographics of compact planetary systems around M dwarf stars. In the
era of TESS, many of these systems are discovered by virtue of the
inner planet transiting the host star, and can often lead to ambiguity
as to the orbital alignment of other planets detected via the RV
technique. Here, we have shown to high statistical significance
($\sim$$2.8\sigma$ and $>20\sigma$ for grazing and central transits,
respectively) that planets c and d do not transit the host star,
raising speculation as to what their true masses may be and if they
are terrestrial in nature. Since the overwhelming majority of all
planets do not transit from a given vantage point, fully
characterizing the bulk of the exoplanet population continues to pose
a challenge for exoplanet demographic studies.

GJ~357~d lies within the CHZ of the host star, resulting in the need
for understanding the nature of planet d to properly assess the
potential for habitable environments within the system. Our dynamical
simulations have shown that the most benign architecture scenario,
where the true mass of planet d is approximately equivalent to the
minimum mass and the orbit is circular, results in 20\% of the HZ
being unstable for other planets in the system. Even a relatively
small eccentricity of 0.1 has the capacity to dramatically increase
regions of instability within the HZ, and many of those cases where
the injected planet survives our simulations result in chaotic orbits
that are unlikely to maintain long-term stability. Therefore, the
widths of the instability regions within the HZ are considered lower
limits on the potentially chaotic influence of planet d. This means
that, though not impossible, it becomes an increasinginly difficult
scenario for the system to harbor an additional Earth-mass planet
within the HZ as the mass and eccentricity of planet d diverge from
their measured lower limits.

Though planet d and its surrounding region may be inhospitable, the
GJ~357 system still has much to offer in the study of planetary
habitability and evolution. For example, planets b and c may be
exceptional candidates for the study of terrestrial planetary
evolution in the high-flux regime of M dwarf stars. Such planets may
be analogous to Venus in their evolution, which can retain significant
volatiles within a post runaway greenhouse atmosphere
\citep{kane2019d,way2020,krissansentotton2021c,garvin2022}. Thus, the
GJ~357 may be an excellent example system to study the boundaries of
planetary habitability, refining target selection approaches to
narrowing the search for possible habitable worlds.


\section*{Acknowledgements}

This research has made use of the Habitable Zone Gallery at
hzgallery.org. The results reported herein benefited from
collaborations and/or information exchange within NASA's Nexus for
Exoplanet System Science (NExSS) research coordination network
sponsored by NASA's Science Mission Directorate.


\software{Mercury \citep{chambers1999}}




\end{document}